\documentclass[
showkeys,	
reprint,	
amsmath,	
amssymb,	
amsfonts,	
aps,		
pre,%
groupedaddress,
superscriptaddress,
a4paper,%
nofootinbib,
eprint,		
final,		
preprintnumbers, 
]{revtex4-2}

\usepackage[tbtags]{mathtools}
\usepackage{amsmath,amsfonts,mathdots,amssymb,yfonts,calc}
\usepackage{graphicx,graphics,xcolor}
\usepackage{subfigure}

\usepackage{natbib}
\usepackage{booktabs}
\usepackage{siunitx}
\usepackage{float}
\usepackage{tabularray}
\usepackage{multirow}
\usepackage{physics}
\usepackage{dcolumn}
\usepackage[utf8]{inputenc}  
\usepackage[T1]{fontenc} 
\usepackage{natbib}
\usepackage{booktabs}
\usepackage{hyperref}
\hypersetup{
    colorlinks=true,
    linkcolor=blue,
    filecolor=magenta,      
    urlcolor=blue!80,
    citecolor=cyan,
    linktocpage=true,
    breaklinks=false,
}

\begin{document}
\title{Universal Curvature Force on Dislocations from a Cartan–Geometric Defect Action}
\author{Vinesh Vijayan}
\email[]{vinesh.physics@rathinam.in}
\affiliation{Department of Science and Humanities, Rathinam Technical Campus, Coimbatore, India -641021}
\author{T. Ishwarya}
\affiliation{Department of Science and Humanities, Rathinam Technical Campus, Coimbatore, India -641021}
\author{M. Parveenbanu}
\affiliation{Department of Science and Humanities, Rathinam Technical Campus, Coimbatore, India -641021}
\author{M Vigneshwaran}
\affiliation{Department of Science and Humanities, Rathinam Technical Campus, Coimbatore, India -641021}

\date{\today}

\begin{abstract}
We develop a unified Cartan–geometric framework where dislocations and disclinations correspond to torsion and curvature of the material coframe connection, respectively, and phase defects emerge as U(1) vortices. This single action principle produces coupled equations of motion and conservation laws governing these defects. Our theory predicts a universal Magnus-like force exerted by curvature on moving dislocations, as well as disclination-driven reconnection events. These phenomena offer experimentally testable signatures in colloidal crystals and mechanical metamaterials.
\end{abstract}

\maketitle
\pagestyle{plain}

\section{\label{s1}Introduction}
Defects dictate the structural, mechanical and transport properties of ordered materials. Dislocations control yielding and plastic deformation; disclinations determine grain-boundary behavior and rotational frustration; vortices dominate the flow of superfluids and superconductors and the turbulence of active nematics. Despite ubiquity across condensed matter, defect types are commonly treated by unrelated theoretical frameworks, and a unifying first-principles theory that predicts \textit{dynamics and interactions} has remained elusive.

Mapping a material containing defects onto a Euclidean reference configuration via a diffeomorphism is fundamentally obstructed by translational and rotational incompatibilities. The nonlinear defect kinematics framework developed by Katanaev and Volovich~\cite{Katanaev1992, Katanaev2005} provides a rigorous geometric theory, characterizing translational incompatibility (dislocations) as torsional densities and rotational incompatibility (disclinations) as curvature densities associated with an underlying connection. Building on this foundation, the nonlinear Riemann-Cartan formulation by Yavari and Goriely~\cite{Yavari2012}, along with comprehensive reviews by Kleman and Friedel~\cite{Kleman2008} and Fressengeas \textit{et al.}~\cite{Fressengeas2017}, have advanced the understanding of defect kinematics, incompatibility, and stress generation in solids. Nonetheless, the derivation of defect forces, reconnection mechanisms, and the evolution of defect networks continues to depend largely on phenomenological models rather than being grounded in an underlying variational principle.

Broader physical roles of defects have emerged from parallel studies in topological condensed matter and mechanical metamaterials. In graphene, dislocations act as sources of torsion, modifying its electronic structure~\cite{DeJuan2010}. In crystalline insulators, dislocation zero modes are often localized at disclinations~\cite{Ruegg2013}, while space-group topology has been systematically correlated with crystalline defects by Slager \textit{et al.}~\cite{Slager2013}. Defect-engineered structures have been demonstrated to host multiple and higher-order topological phases~\cite{Benalcazar2017}, and orientational defects in mechanical lattices have been shown to activate topological vibrational modes~\cite{Liang2021}. In a study of mechanical metamaterials, tailoring defect geometries can induce robust floppy modes and shape-changing mechanisms, revealing the interplay between topology and mechanics~\cite{MaoLubensky2019}. Defect dynamics have been implicated in driving complex phenomena such as flows, chaos, and turbulence in active and soft matter systems~\cite{Shankar2018, Doostmohammadi2018, Giomi2015}, with recent studies quantifying defect kinetics and interactions in active nematics~\cite{LemmaShankarBowick2021}. Collectively, these studies highlight that defects function as intrinsic geometric entities whose dynamics and interactions extend beyond the conventional framework of elasticity.

Building on these observations, we present a Cartan-geometric defect action that unifies translational, rotational, and phase defects within a single cohesive framework. Utilizing Cartan's generalization of Riemannian geometry~\cite{Cartan1922,Cartan1923,Cartan1925,Schrodinger1950,Trautman2006}, the coframe $e^a$ and spin connection $\omega^{ab}$ encode the underlying crystallographic structure. Interpreting torsion and curvature as dislocation and disclination densities respectively, the variation of the Riemann-Cartan action produces force and coupled-stress balance equations analogous to classical gauge-theoretic treatments~\cite{Hehl1976, HehlObukhov2007, HehlObukhov2012}. The resulting geometric field equations yield canonical analytical solutions for screw and edge dislocations as well as wedge dislocations, in agreement with foundational geometric defect models~\cite{Kondo1952, Edelen1983, Malyshev2000, Mikhail1977}.

Most importantly, the theory predicts a new, experimentally accessible effect: a curvature-induced transverse force acting on moving dislocations. We further show that disclinations mediate the reconnection and annihilation of dislocation lines, allowing the framework to be extended to encompass U(1) phase defects and thereby providing a unified description of lattice defects and vortices.

The paper is organized as follows. Section~\ref{s2} summarizes the geometric field concepts and variational action and the dynamical equations; Section~\ref{s3} derives the defect solutions; Section~\ref{s4} derives the universal curvature force;Section~\ref{s5}  analyzes curvature-induced Burgers vector exchange and dislocation reconnections; Section~\ref{s6} introduces the phase factor and vortex flow dynamics, followed by discussion in Section~\ref{s7} and conclusions in Section~\ref{s8}.
\section{Cartan--Geometry and Geometric Action}
\label{s2}
Let \(\mathcal{M}\) be a \(d\)-dimensional oriented manifold modeling a continuous solid. Consider a set of 1-forms 
\begin{equation}
e^a = e^a_{\ \mu} dx^\mu, \quad a=1,\ldots,n,
\label{E1}
\end{equation}
which form a local basis of the cotangent space at each point. The 1-forms \(e^a\) encode the local crystallographic frame or elastic distortion of the material. Integrating \(e^a\) along a curve yields the material displacement in the internal directions.

\begin{figure}[t]
  \centering
  \includegraphics[width=0.85\columnwidth]{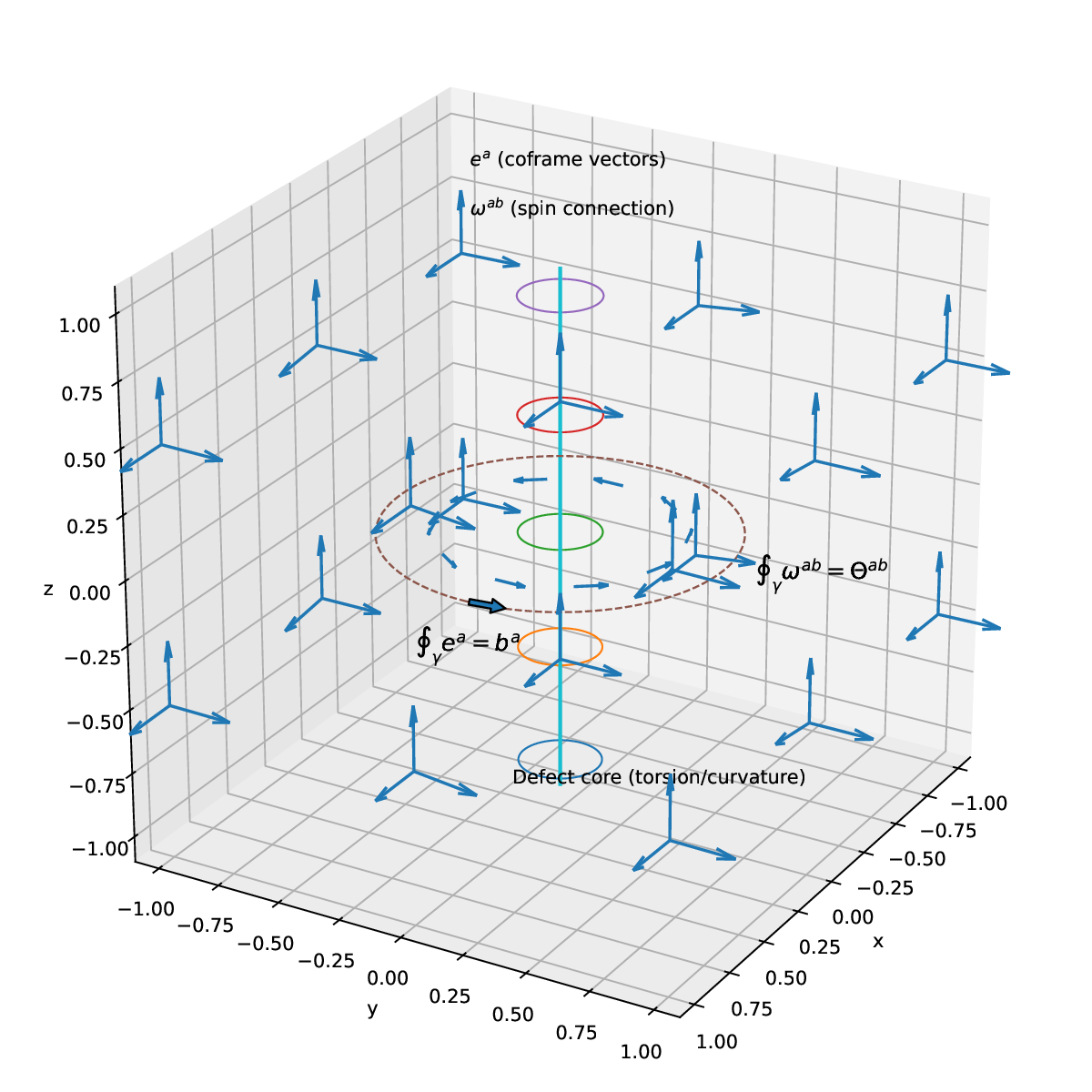}
  \caption{Schematic geometry of coframe \(e^a\) and spin connection \(\omega^{ab}\).}
  \label{fig1}
\end{figure}

Let \(\omega^a_{\ b}\) be an \(\mathfrak{so}(n)\)-valued 1-form,
\begin{equation}
\omega^a_{\ b} = \omega^a_{\ b \mu} dx^\mu, \quad \omega_{ab} = - \omega_{ba},
\label{E2}
\end{equation}
which encodes how the local frame is rotated from point to point (an $\mathrm{SO}(3)$ connection $\omega^a{}_b$)[see Figure~\ref{fig1}].
The covariant derivative of \(e^a\), also known as the torsion 2-form, is defined as
\begin{equation}
T^a = D e^a = d e^a + \omega^a_{\ b} \wedge e^b,
\label{E3}
\end{equation}
which measures defects in the translational order of the lattice (dislocations). A nonzero value of \(T^a\) means small parallelograms fail to close in the real internal frame.

Similarly, the curvature 2-form \(R^a_{\ b}\) measures defects in the rotational order (disclinations) and is defined as
\begin{equation}
R^a_{\ b} = D \omega^a_{\ b} = d \omega^a_{\ b} + \omega^a_{\ c} \wedge \omega^c_{\ b}.
\label{E4}
\end{equation}
The field strengths \(T^a\) and \(R^a_{\ b}\) represent the densities of dislocations and disclinations, respectively. Specifically, \(T^a\) measures the translational defects, while \(R^a_{\ b}\) encodes the rotational defects in the material.
A dislocation is a line defect in a crystal where parallel transport in space does not quite add up translationally. In ordinary elasticity, if we take a closed loop around the defect consisting of small lattice steps, when returning to the starting point in the lab coordinates, we will not land where expected according to a perfect lattice. This mismatch is known as the Burgers vector.

Torsion measures the failure of infinitesimal parallelograms to close under parallel transport, characterizing dislocations. In Cartan geometry, the torsion flux through a surface equals the total Burgers vector:
\begin{equation}
\int_S T^a \sim b^a,
\label{E5}
\end{equation}
so dislocations act as torsion sources.

Disclinations are rotational defects causing a local frame to rotate by the Frank angle \(\Omega\) after parallel transport around a loop, defining the Frank vector. The curvature flux through a surface equals the total Frank vector:
\begin{equation}
\int_S R^a_{\ b} \sim \Omega^a_{\ b},
\label{E6}
\end{equation}
making disclinations sources of curvature.

Holonomy—the change from parallel transport around a closed loop—has translational and rotational parts given by \eqref{E5} and \eqref{E6}. Thus, Burgers and Frank vectors are topological charges: holonomies of the Cartan connection gauge field, not just elastic parameters.

Cartan geometry satisfies geometric identities closely related to the Bianchi identities, which directly lead to conservation laws:
\begin{equation}
\begin{aligned}
D R^{ab} &= 0, \\
D T^a  &= R^a_{\ b} \wedge e^b.
\end{aligned}
\label{E7}
\end{equation}
The first identity forbids isolated endpoints of disclination lines, requiring that they either form closed loops or terminate at boundaries. The second identity states that the change of dislocation density is governed by curvature; that is, disclinations act as sources of dislocations. These represent the covariant conservation laws for defect currents, governing the dynamics of defects.

Thus, the allowed worldlines and worldsheets of defects are governed by the geometric identities of the Cartan connection. No ad hoc conservation laws need be imposed; they arise naturally from the geometry.

\subsection{The Defect Action}
We postulate an action constructed from \(e^a\) (the coframe representing elastic degrees of freedom), \(\omega^a_{\ b}\) (the spin connection representing the orientational field), torsion, and curvature. This action defines a differential geometric field theory of material defects, where torsion and curvature interact dynamically, unifying defect dynamics within a gauge geometry of the Euclidean group. The defect action
\begin{equation}
S[e,\omega] = \int \big[ \alpha\, T^a \wedge *T_a 
+ \beta\, R^a{}_b \wedge *R^b{}_a
+ \gamma\, e^a \wedge R_{ab} \wedge e^b \big],
\label{E8}
\end{equation}
with moduli $\alpha,\beta,\gamma > 0$.

The first term in the action represents the elastic dislocation energy, comprising both the core energy and the long-range elastic energy of dislocations. Mathematically, it is an exact analogue of the Yang-Mills term and captures the elastic stiffness associated with shear and Burgers vector distortions. The second term accounts for the orientational defect energy—the energetic cost of curvature—providing the material's resistance to rotational distortions, analogous to the Frank energy in liquid crystals. The third term is a crucial new contribution: a mixed torsion-curvature elasticity. This natural 4-form plays a purely condensed matter role, expressing how rotational defects distort the geometry defined by \(e^a\). Variation of this term produces Magnus-like forces. Importantly, it allows dynamic interaction between torsion and curvature, elevating the theory beyond conventional elasticity into a gauge theory of the Euclidean group. Analogous to the Einstein-Hilbert term in gravitational theory, this term couples the fermionic field to spacetime curvature in materials with coupled translational and rotational stiffness. It is the simplest invariant that encodes how the presence of a disclination locally modifies the effective metric that dislocations respond to.

The Euler–Lagrange equations are derived as follows. The covariant variations of equations \eqref{E3} and \eqref{E4} are
\begin{equation}
\begin{aligned}
\delta T^a &= D(\delta e^a) + \delta \omega^a_{\ b} \wedge e^b, \\
\delta R^a_{\ b} &= D(\delta \omega^a_{\ b}).
\end{aligned}
\label{E9}
\end{equation}
We assume that the Hodge star operator has no variation, as the background metric—defined independently by \(e^a\)—is fixed. Boundary terms are dropped by imposing standard physical boundary conditions.
\subsubsection{Variation with respect to co-frame $e^a$}
The action terms that depend on \(e^a\) are the torsion-squared term and the mixed torsion-curvature term, while the curvature-squared term is independent of \(e^a\).

The variation of the torsion term yields
\begin{equation}
\delta S_T = \alpha \int \left[ \delta T^a \wedge * T_a + T^a \wedge \delta(* T_a) \right].
\label{E10}
\end{equation}
Since \(\delta(* T_a) = *(\delta T_a)\), the variation simplifies accordingly. Equation(\ref{E10}) now becomes
\begin{equation}
\delta S_T =2\alpha \int \delta T^a \wedge *T_a
\label{E11}
\end{equation}
substituting Equation(\ref{E9})in to Equation (\ref{E11})
\begin{equation*}
\delta S_T =2\alpha \int[D(\delta e^a) + \delta \omega^a_b \wedge e^b] \wedge *T_a
\end{equation*}
The second term belongs to the $\omega$ equation. So the contribution from the torsion term is obtained by integrating by parts the first term  and neglecting the boundary term as follows
\begin{equation}
\delta S_T =2\alpha \int D(\delta e^a)\wedge *T_a =-2 \alpha \int \delta e^a \wedge D(*T_a)
\label{E12}
\end{equation}
The curvature dependent term is independent of $e^a$ variations and hence
\begin{equation}
\delta S_C= 0
\label{E13}
\end{equation}
The variation of the mixed term is given by
\begin{equation}
\delta s_M = \gamma \int \delta e^a \wedge R_{ab} \wedge e^b + \gamma \int e^a \wedge R_{ab} \wedge \delta e^b.
\label{E14}
\end{equation}

Swapping the dummy indices \(a \leftrightarrow b\) in the last term and combining both contributions yields
\begin{equation}
\delta s_M = 2 \gamma \int \delta e^a \wedge R_{ab} \wedge e^b.
\label{E15}
\end{equation}

Now, the total variation of the co-frame is
\begin{equation}
\delta s = \int \delta e^a \wedge \left[-2 \alpha\, D(*T_a) + 2 \gamma\, R_{ab} \wedge e^b \right].
\label{E16}
\end{equation}

The corresponding Euler--Lagrange equation, obtained by demanding stationarity with respect to \(\delta e^a\), is
\begin{equation}
D(*T_a) + \Gamma R_{ab} \wedge e^b = 0.
\label{E17}
\end{equation}
where $\Gamma=\frac{\gamma}{\alpha}$.
\subsubsection{Variation with respect to the spin connection $\omega^{ab}$}
From the torsion-squared term, we have
\begin{equation}
\delta S_T = 2 \alpha \int \delta \omega_{ab} \wedge e^b \wedge *T_a.
\label{E18}
\end{equation}
The variation of the curvature-squared term, after integration by parts and neglecting boundary contributions, yields
\begin{equation}
\delta S_C = -2 \beta \int \delta \omega_{ab} \wedge D(*R_{ba}).
\label{E19}
\end{equation}

The variation of the mixed term is
\begin{equation*}
\delta S_{\text{mix}} = \gamma \int e^a \wedge D(\delta \omega_{ab}) \wedge e^b.
\end{equation*}

Integrating by parts, this becomes
\begin{equation}
\delta S_{\text{mix}} = - \gamma \int \delta \omega_{ab} \, D(e^a \wedge e^b).
\label{E20}
\end{equation}

Using the identity
\[
D(e^a \wedge e^b) = T^a \wedge e^b - e^a \wedge T^b,
\]

we obtain
\begin{equation}
\delta S_{\text{mix}} = - \gamma \int \delta \omega_{ab} \left[ T^a \wedge e^b - e^a \wedge T^b \right].
\label{E21}
\end{equation}

Collecting all \(\omega\) variations from Equations \eqref{E18}, \eqref{E19}, and \eqref{E21}, we have
\begin{align}
\delta \omega_S = \int \delta \omega_{ab} \Big[\, &2 \alpha\, e^b \wedge *T_a - 2 \beta\, D(*R_{ab}) \notag \\
&- \gamma \left( T^a \wedge e^b - e^a \wedge T^b \right) \Big].
\label{E22}
\end{align}
The corresponding Euler–Lagrange equation is
\begin{equation}
D(*R_{ab}) + \kappa \left( e^a \wedge *T_b - e^b \wedge *T_a \right) = 0,
\label{23}
\end{equation}
where \(\kappa = \frac{\gamma}{2 \beta}\).

The action given in Equation~\eqref{E8} leads to two geometric field equations: one describing the force balance, where dislocations are driven by curvature [Equation (\ref{E17})], and the other describing the spin moment balance, where curvature responds to torsion [Equation (\ref{23})].

Dislocation and disclination transport follow from the Bianchi identities, Equation~\eqref{E7}. These equations provide the dynamics \textit{without constitutive postulates}.

\section{Canonical Defect Solutions}
\label{s3}
We show that Eq.~\eqref{E17} and Eq.~\eqref{23} reproduce the classical defect fields. In what follows, we work in 3D space with coordinates \((x,y,z)\), and use polar coordinates \((r,\theta,z)\) in the \(xy\)-plane when needed.

\subsection{Torsion for a screw dislocation}
Set up a screw dislocation along the \(z\)-axis with Burgers vector \(\mathbf{b} = b\, \hat{z}\). Choose the coframe
\begin{equation}
e^1 = dx, \quad e^2 = dy, \quad e^3 = dz + \frac{b}{2\pi} d\theta,
\label{E24}
\end{equation}
where going around the origin \(\theta: r \to \theta + 2\pi\), the \(z\)-coordinate jumps by \(b\), which describes the classic screw dislocation.

Assuming a pure dislocation, the torsion from Eq.~(\ref{E3}) is given by
\begin{equation}
T^a = d e^a.
\label{E25}
\end{equation}

For \(a=1,2\) and \(e^1 = dx, e^2 = dy\),
\begin{equation}
T^1 = d(dx) = 0, \quad T^2 = d(dy) = 0.
\label{E26}
\end{equation}

For \(a=3\),
\begin{equation}
T^3 = d e^3 = d\left( dz + \frac{b}{2\pi} d\theta \right).
\label{E27}
\end{equation}

Classically, \(d(d\theta) = 0\), but since \(\theta\) is multivalued, it must be treated as a distribution.

The distributional identity is
\begin{equation}
d(d\theta) = 2\pi \delta^{(2)}(r) \, dx \wedge dy,
\label{E28}
\end{equation}
where in the plane for \(r > 0\),
\[
\theta = \tan^{-1}\left( \frac{y}{x} \right), \quad
d\theta = \frac{-y\, dx + x\, dy}{x^2 + y^2} = \frac{-y\, dx + x\, dy}{r^2}.
\]

Away from the origin, this is a smooth one-form and \(d(d\theta) = 0\).

Integrating around the origin over a disk \(D\) gives
\begin{equation}
\int_D d(d\theta) = \oint_{\partial D} d\theta = \int_0^{2\pi} d\theta = 2\pi.
\label{E29}
\end{equation}
\begin{figure*}[htbp]
  \centering
  \includegraphics[width=0.85\linewidth]{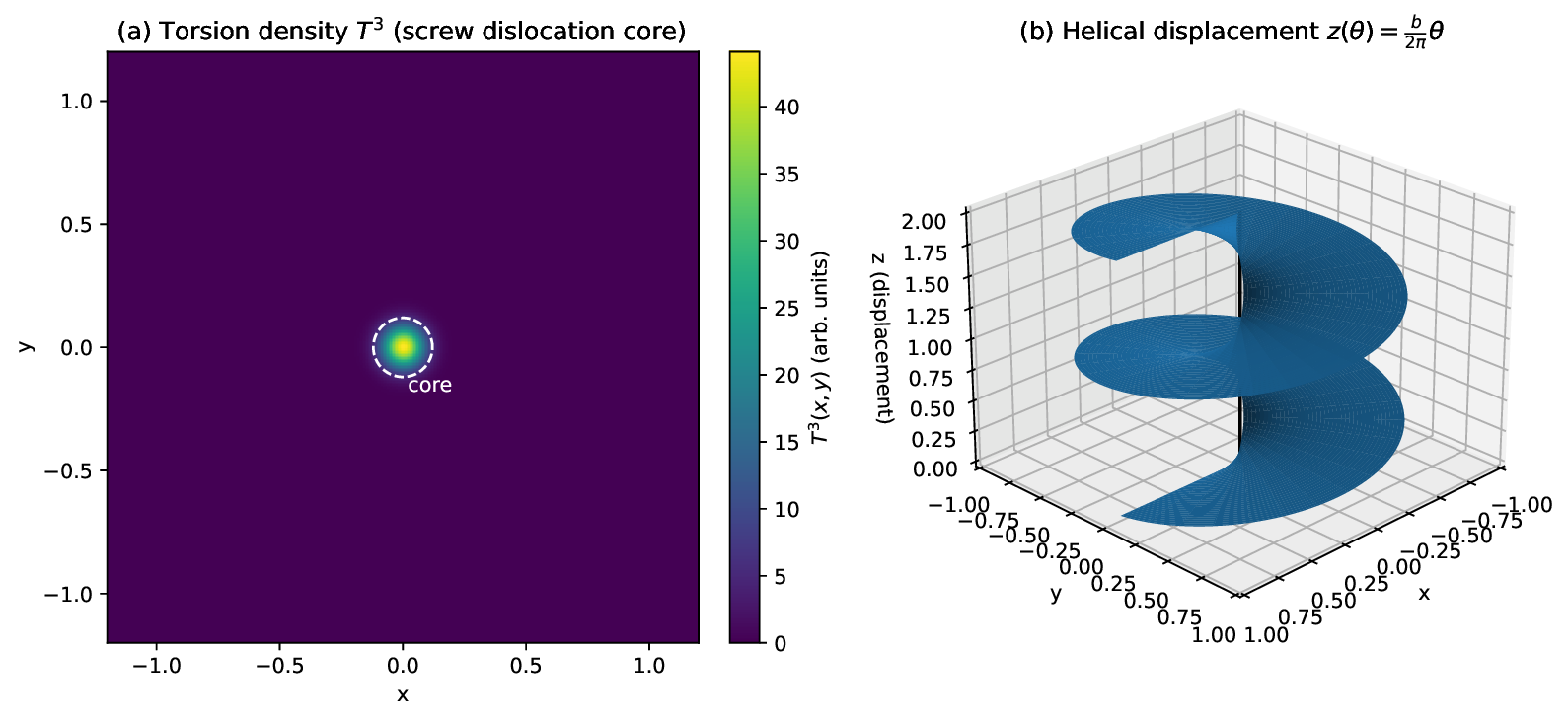}
  \caption{
    \textbf{(a)} Torsion density $T^3(x,y)$ for a screw dislocation, approximated by a narrow Gaussian to visualize the core; the analytic result is $T^3=b\,\delta^{(2)}(r)\,dx\wedge dy$. 
    \textbf{(b)} Helical displacement illustrating $e^3=dz+\tfrac{b}{2\pi}d\theta$ and $z(\theta)=\tfrac{b}{2\pi}\theta$ (two full turns shown). 
    The central axis denotes the defect line.
  }
  \label{fig2}
\end{figure*}
Therefore, from Eq.~\eqref{E27} and Eq.~\eqref{E28},
\begin{equation}
T^3 = \frac{b}{2\pi} d(d\theta) = \frac{b}{2\pi} \cdot 2\pi \delta^{(2)}(r) \, dx \wedge dy = b\, \delta^{(2)}(r) \, dx \wedge dy,
\label{E30}
\end{equation}
so the torsion is localized along the \(z\)-axis.

Hence, the screw dislocation corresponds to a line of torsion flux concentrated at the origin, with
\begin{equation}
T^3 = b\, \delta^{(2)}(r) \, dx \wedge dy, \quad T^1 = T^2 = 0.
\label{E31}
\end{equation}
Thus, the geometric structure encodes the screw dislocation exactly.The scenario is shown in Figure~\ref{fig2}.

\subsection{Edge dislocation}
Consider an edge dislocation with Burgers vector \(\mathbf{b} = b \hat{x}\). The coframe is defined as
\begin{equation}
\begin{aligned}
e^1 &= dx + \frac{b}{2\pi} \frac{y}{x^2 + y^2} d\theta, \\
e^2 &= dy - \frac{b}{2\pi} \frac{x}{x^2 + y^2} d\theta, \\
e^3 &= dz.
\end{aligned}
\label{E32}
\end{equation}

This coframe induces an in-plane displacement field without rotation. The torsion 2-forms are
\begin{equation}
T^1 = b\, \delta^{(2)}(r)\, dy \wedge dz, \quad
T^2 = -b\, \delta^{(2)}(r)\, dx \wedge dz,
\label{E33}
\end{equation}
with \(T^3 = 0\) and \(\omega^a{}_b = 0\).

\begin{figure*}[t]
\centering
\includegraphics[width=0.85\linewidth]{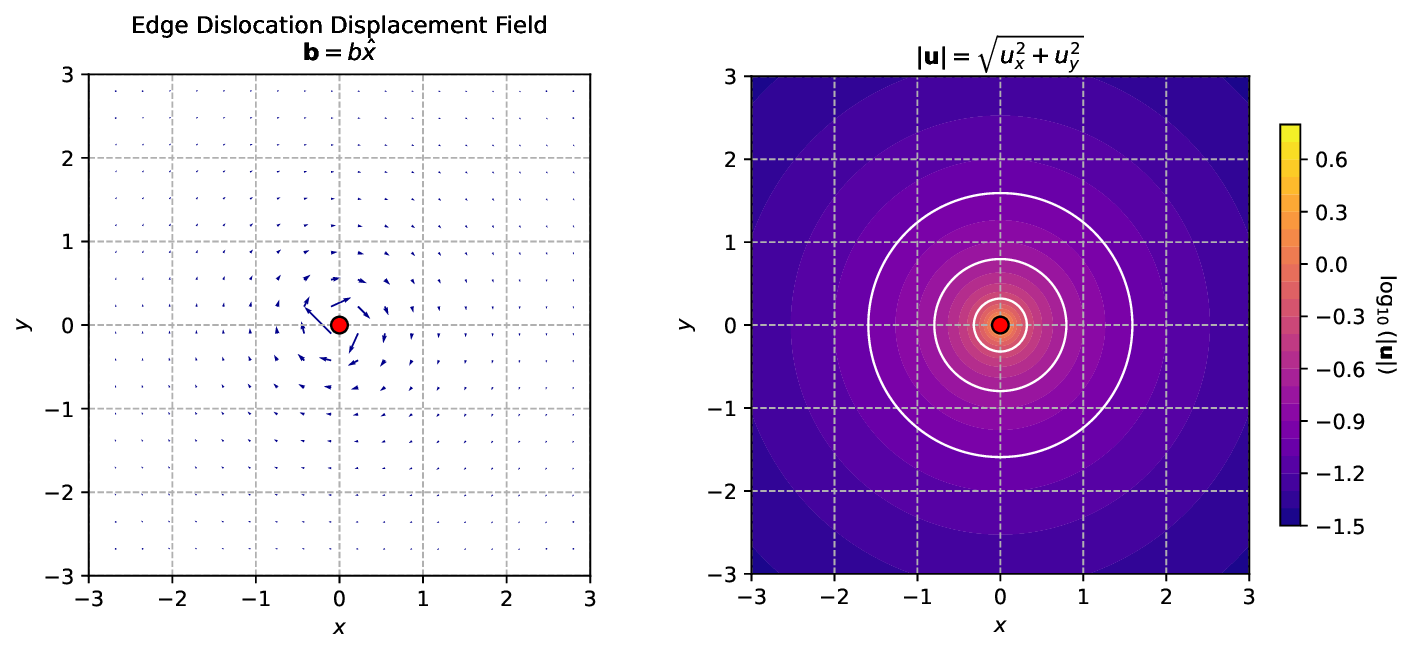}
\caption{
\textbf{Geometric displacement field of edge dislocation.} 
Quiver plot shows the in-plane displacement field $\mathbf{u} = (u_x, u_y)$ 
from the coframe Eq.~\eqref{E32}. 
The red dot marks the dislocation core at the origin. 
The field exhibits characteristic $1/r$ decay with circulation corresponding to Burgers vector $\mathbf{b} = b\hat{x}$.
}
\label{fig3}
\end{figure*}

Thus, the defect is characterized by pure torsion with zero curvature. Unlike the screw dislocation, the Burgers vector for the edge dislocation lies entirely in the \(xy\)-plane. The displacement caused by the defect involves a shift within the plane rather than a helical twist around the axis. Figure~\ref{fig3} shows the geometric displacement field of edge dislocation.

\subsection{Wedge Disclination}
\begin{figure}[htbp]
\centering
\includegraphics[width=0.9\columnwidth]{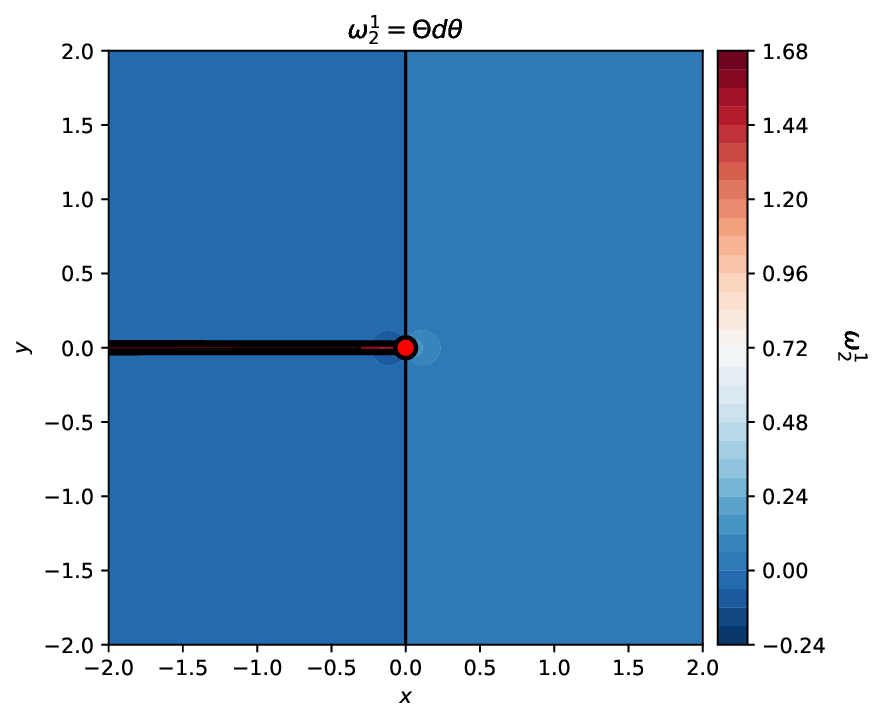}
\caption{
Connection $\omega^1_2 = \Theta d\theta$ for wedge disclination. The circulatory field around the core (red dot) produces curvature $R^1_2 = 2\pi \Theta \delta^{(2)}(r)$ upon differentiation.
}
\label{fig4}
\end{figure}

A wedge disclination with Frank angle $\Theta$ along the $z$-axis arises from a rotational mismatch. Using the trivial coframe
\begin{equation}
e^1 = dx, \quad e^2 = dy, \quad e^3 = dz,
\label{E34}
\end{equation}
and a non-trivial connection encoding rotation in the $xy$-plane,
\begin{equation}
\omega^1{}_2 = \Theta\, d\theta, \quad \omega^2{}_1 = -\Theta\, d\theta, \quad \omega^1{}_3 = \omega^2{}_3 = 0,
\label{E35}
\end{equation}
the curvature is
\begin{equation}
R^1{}_2 = d\omega^1{}_2 = 2\pi \Theta\, \delta^{(2)}(r)\, dx \wedge dy,
\label{E36}
\end{equation}
with all other $R^a{}_b = 0$. Thus, the disclination is characterized by pure curvature with zero torsion. The integral of the curvature over any surface piercing the line gives the Frank rotation: $\int_S R^1{}_2 = 2\pi \Theta$. This scenario is illustrated in Figure~\ref{fig4}.

\section{Universal Curvature Force on Moving Dislocations}\label{s4}

Applying the Bianchi identity to the transport of torsion gives
\begin{equation}
\partial_t(*T_a) + \mathcal{L}_v(*T_a) = D(\iota_v *T_a) + \iota_v D(*T_a),
\label{37}
\end{equation}
where \(\mathcal{L}_v\) is the Lie derivative along the dislocation velocity \(\mathbf{v}\) and \(\iota_v\) is the interior product. Substituting the field equation Eq.~\eqref{E17} into this relation yields
\begin{equation}
\partial_t(*T_a) + \mathcal{L}_v(*T_a) = D(\iota_v *T_a) - \Gamma\, \iota_v \big( R_{ab} \wedge e^b \big).
\label{38}
\end{equation}
The left-hand side represents the material time derivative of the torsion flux, i.e., the evolution of the dislocation line. The first term on the right-hand side corresponds to an inertial transport term, while the second term encodes the configurational force exerted by curvature.

For a screw dislocation moving with velocity $\mathbf{v} = v_\perp \hat{x}$, the torsion forms a flux tube concentrated along the core: 
$T^3 = b \delta^{(2)}(r_\perp) dx \wedge dy$, 
where $\mathbf{b} = b \hat{z}$ is the Burgers vector parallel to the core direction $\hat{\mathbf{t}} = \hat{z}$. This perpendicular motion $v_\perp$ transports the torsion flux through the surrounding geometry. 
The Hodge dual $*T_3 \sim b \delta^{(2)}(r_\perp) \hat{z}$ represents the flux density along the core direction, quantifying the dislocation strength as a localized translational defect line. This geometric structure directly couples to the ambient curvature field through the field equations.

For a wedge disclination along the $z$-axis with Frank vector $\boldsymbol{\Theta} = \Theta \hat{z}$, the curvature is
\begin{equation}
R^1{}_2 = 2\pi \Theta \delta^{(2)}(r)\, dx \wedge dy, \quad R_{3b} \wedge e^b \sim \Theta \delta^{(2)}(r)\, \hat{z}.
\label{E39}
\end{equation}

Projecting the transport equation onto the core tangent $\hat{\mathbf{t}}$ gives
\begin{equation}
\hat{\mathbf{t}} \cdot \big[ \partial_t (*T_a) + \mathcal{L}_v (*T_a) \big]
= - \gamma\, \hat{\mathbf{t}} \cdot \iota_v (R_{ab} \wedge e^b).
\label{E40}
\end{equation}

The left-hand side represents flux transport: $\hat{z} \cdot \mathcal{L}_v (*T_3) \sim b v_\perp / A$. The key right-hand side calculation is
\begin{equation}
\iota_v (R_{3b} \wedge e^b) = \iota_v \big( \Theta \delta^{(2)}(r)\, \hat{z} \big)
= \Theta \delta^{(2)}(r)\, (\mathbf{v} \times \hat{z}),
\label{E41}
\end{equation}
using the interior product rule $\iota_v(\alpha \wedge \beta) = (\iota_v \alpha)\beta - (-1)^{\deg \alpha} \alpha \wedge (\iota_v \beta)$. Thus,
\begin{equation}
\hat{z} \cdot \iota_v (R_{3b} \wedge e^b) = \Theta \delta^{(2)}(r)\, \big[ \hat{z} \cdot (\mathbf{v} \times \hat{z}) \big]
= \Theta v_\perp,
\label{E42}
\end{equation}
yielding the transverse configurational force
\begin{equation}
F_\perp = \Gamma\, (\boldsymbol{\Theta} \times \mathbf{b}) \times v.
\label{E43}
\end{equation}

This means a moving dislocation in a curvature field experiences a transverse force perpendicular to both velocity and Frank vector---exactly like a charged particle in a magnetic field or a spinning cylinder in a fluid. Thus, the action generates universal coupling between translational and rotational defects, predicting this effect without material constants. This can be interpreted as a topological Hall effect for defects, where curvature acts as an effective magnetic field deflecting moving dislocations. 
  
\begin{figure}[htbp]
\centering
\includegraphics[width=0.85\columnwidth]{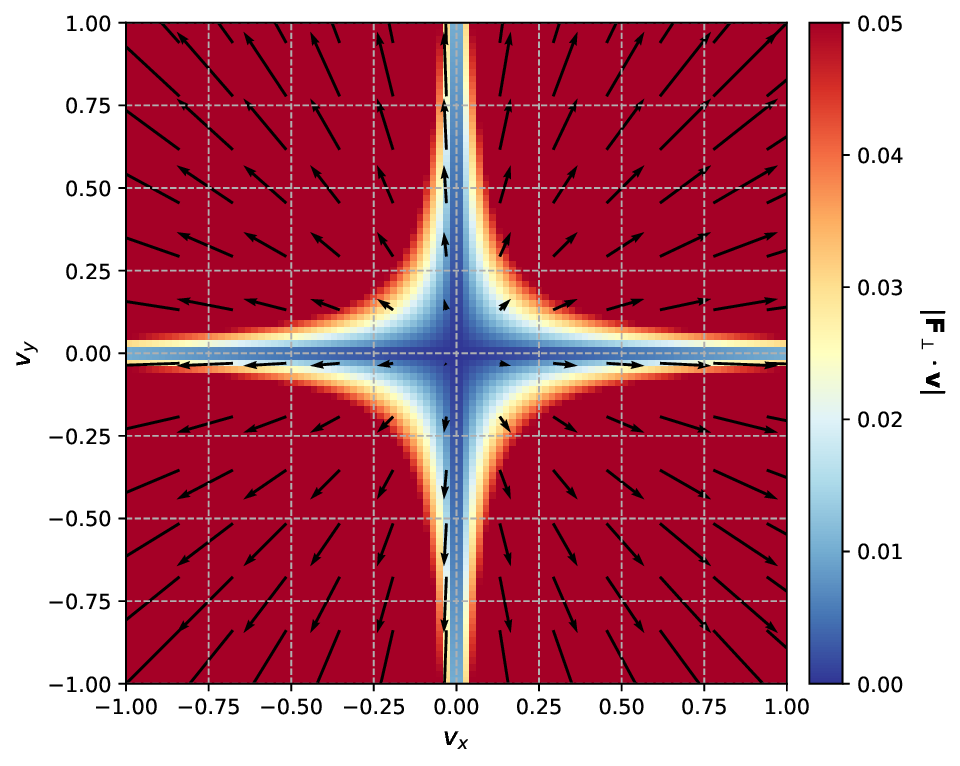}
\caption{
\textbf{Magnus-like configurational force on a moving dislocation.} The color map shows the magnitude of the scalar product $|\mathbf{F}_\perp \cdot \mathbf{v}|$ of the Magnus-type force and velocity. The thick blue contour marks the locus where the scalar product is exactly zero, demonstrating that the configurational force is strictly transverse to the dislocation velocity for all directions in the slip plane. Black arrows indicate sample velocity directions $\mathbf{v}$ for reference.
}
\label{fig5}
\end{figure}

Eq~\eqref{E43} is the central new prediction of this work: \textbf{curvature generates a universal Magnus-like transverse force on moving dislocations, independent of elastic moduli}. This effect has not previously appeared in elasticity, micropolar mechanics or geometric formulations. Figure~\ref{fig5} depicts Magnus-like configurational force on a moving dislocation.

\section{Reconnection and Annihilation of Dislocation Lines}\label{s5}
Integrating the Bianchi identity over any oriented 3-volume $V$,
\begin{equation}
\int_V D T^a = \int_V R^a_{\ b} \wedge e^b,
\label{E44}
\end{equation}
and applying Stokes' theorem for the covariant exterior derivative,
\begin{equation*}
\int_V D T^a = \int_{\partial V} T^a,
\end{equation*}
yields
\begin{equation}
\int_{\partial V} T^a = \int_V R^a_{\ b} \wedge e^b.
\label{E45}
\end{equation}

The left-hand side represents the \textbf{net torsion (Burgers flux)} escaping through the boundary $\partial V$. When $\partial V$ is chosen to intersect incoming/outgoing dislocations, $\int_{\partial V} T^a$ equals the \textbf{algebraic sum of Burgers vectors} piercing the surface. The right-hand side is the \textbf{curvature 2-form source} integrated over $V$. Thus Eq.~\eqref{E45} is the \textbf{exact topological statement} that disclinations (curvature) act as sources or sinks of Burgers flux (dislocations).

Choose the volume $V$ surrounding the reconnection event, with surface $\partial V$ as the union of small cross-sections $S_1^-, S_2^-, \dots, S_n^-$ (incoming) and $S_1^+, S_2^+, \dots, S_m^+$ (outgoing). Orient all cross-sections such that the \textbf{incoming fluxes are positive}. Then Eq.~\eqref{E45} gives
\begin{equation}
\sum_{\rm in} b_{\rm in}^a - \sum_{\rm out} b_{\rm out}^a = \int_V R^a_{\ b} \wedge e^b.
\label{E46}
\end{equation}

For the simplest reconnection event---two incoming dislocations merging into one outgoing line:
\begin{equation}
b_1^a + b_2^a - b_f^a = \int_V R^a_{\ b} \wedge e^b,
\label{E47}
\end{equation}
which rearranges to
\begin{equation}
b_1^a + b_2^a + \Delta b^a(\boldsymbol{\Theta}) = 0,
\label{E48}
\end{equation}
where $\Delta b^a(\boldsymbol{\Theta}) = -\int_V R^a_{\ b} \wedge e^b$ is the \textbf{curvature-screened Burgers flux}.

For \textbf{annihilation} ($b_f^a = 0$), $\Delta b^a(\boldsymbol{\Theta}) = -(b_1^a + b_2^a)$, so the disclination absorbs the net Burgers charge. The sign of $\Delta b^a$ is determined by the curvature $\boldsymbol{\Theta}$, selecting attraction/repulsion and reaction channels.

\section{Coupling to U(1) Phase Defects}\label{s6}

To treat vortex-like defects on equal footing with lattice defects, we introduce a U(1) gauge field $A$ with $F=dA$ and add
\begin{equation}
S_{\text{phase}} = \kappa \int T^a \wedge F \wedge e_a 
+ \lambda \int e^a \wedge R_{ab} \wedge F \wedge e^b.
\label{E49}
\end{equation}
The variation of $S_{\rm phase}$ with respect to the gauge potential $A$ gives
\begin{equation}
\delta_A S_{\rm phase} = \kappa \int_M T^a \wedge \delta F \wedge e_a + \lambda \int_M e^a \wedge R_{ab} \wedge \delta F \wedge e^b,
\label{E50}
\end{equation}
where $\delta F = d(\delta A)$. Substituting yields
\begin{equation}
\delta_A S_{\rm phase} = \kappa \int_M T^a \wedge d(\delta A) \wedge e_a + \lambda \int_M e^a \wedge R_{ab} \wedge d(\delta A) \wedge e^b.
\label{E51}
\end{equation}

Applying the Leibniz rule $d(\alpha \wedge \beta) = d\alpha \wedge \beta + (-1)^{\deg \alpha} \alpha \wedge d\beta$ and integrating by parts (with boundary terms vanishing by Stokes' theorem),
\begin{equation}
\delta_A S_{\rm phase} = -\int_M \left[ \kappa \, d(T^a \wedge e_a) + \lambda \, d(e^a \wedge R_{ab} \wedge e^b) \right] \wedge \delta A.
\label{E52}
\end{equation}

The Euler-Lagrange equation for $A$ is thus
\begin{equation}
d(*F) = \kappa \, d(T^a \wedge e_a) + \lambda \, d(e^a \wedge R_{ab} \wedge e^b).
\label{E53}
\end{equation}
By retaining $\delta F = d(\delta A)$ and moving $d$ off $\delta A$ via integration by parts, we obtain
\begin{equation}
d(*F) = \kappa T^a \wedge e_a + \lambda e^a \wedge R_{ab} \wedge e^b
,
\label{E54}
\end{equation}
provided the elementary forms $T^a \wedge e_a$ and $e^a \wedge R_{ab} \wedge e^b$ are exact (absorbing the overall $d$).

The left-hand side $d(*F)$ represents the divergence of the U(1) field strength. Integrating over a volume $V$ and applying Stokes' theorem yields the net U(1) charge (displacement flux) leaving the boundary:
\begin{equation}
\int_{\partial V} *F = \int_V d(*F) = \int_V J,
\label{E55}
\end{equation}
where $J = \kappa T^a \wedge e_a + \lambda e^a \wedge R_{ab} \wedge e^b$.

The right-hand side shows that torsion and curvature supply/remove U(1) flux, so dislocations and disclinations act as sources and sinks of vortex charge. 

Applying $d$ to both sides of Eq.~(\ref{E55}) gives the consistency condition:
\begin{equation}
d(d*F) = dJ = 0,
\label{E56}
\end{equation}
requiring the geometric sources to be closed forms.
Consequences:
(i) vortex pinning on dislocation cores;  
(ii) curvature-assisted vortex conversion;  
(iii) mixed Burgers--vortex topological charge.

\section{Discussion}\label{s7}
The geometric method discussed here builds upon and extends the seminal geometric descriptions of defects. The foundational work by Katanaev and Volovik established a kinematical approach, identifying dislocations and disclinations with torsion and curvature, respectively. The nonlinear Riemann--Cartan formulation by Yavari and Goriely applied this geometry to derive equations of finite elasticity for dislocated bodies within the limit of vanishing curvature (Weitzenb\"ock limit). In our calculation, we proceeded in two key directions. First, we treat torsion and curvature as independent fields from the outset, employing a unified action principle rather than imposing constraints or focusing solely on elasticity. Second, the mixed torsion-curvature term in the action introduces a minimal dynamical coupling between defect sectors. This coupling term plays a crucial role in predicting new physical effects, most notably the Magnus-like transverse force and the explicit disclination-mediated reconnection dynamics discussed in Section~(V). Both results naturally follow from the variational principle without any ad hoc phenomenological input. Thus, the action provides a dynamical and coupled gauge theory of the defect network itself.

The velocity-dependent transverse drift, perpendicular to the dislocation velocity and described by Equation~\ref{E43}, scales with both the Frank angle and the Burgers vector. This transverse force differs from the velocity-independent Peach-Koehler~\cite{PeachKoehler1950, PeachKoehler1953} force and drag forces, which align with the dislocation velocity, as well as from thermal noise or classical elasticity effects. Its emergence arises solely from geometric coupling, which models defects as singular curvature sources at length scales larger than the lattice spacing, thereby capturing their essential behavior in the linear regime for well-separated defects. Although the theory neglects atomic-scale and nonlinear effects, extensions incorporating discrete or stochastic phenomena can be developed without altering the fundamental mechanism.

In systems where both defect motion and curvature are controllable and observable, experimental validation of the curvature-induced transverse force is feasible. Suitable platforms include colloidal crystals, mechanical metamaterials, active nematic films, and strained two-dimensional materials such as graphene. The characteristic velocity-dependent transverse motion, governed by the Burgers vector and Frank angle, can be revealed through advanced microscopy and particle-tracking techniques. Observation of this mechanism will provide compelling evidence for geometric coupling and open new avenues for manipulating defects via curvature control.

\section{Conclusion}\label{s8}

A unified Cartan–geometric theory is constructed in which torsion and curvature represent dislocations and disclinations, respectively. The corresponding action yields the correct analytical solutions and conservation laws, and predicts a universal curvature-induced transverse force on moving dislocations. It further accounts for disclination-mediated reconnection of dislocation lines and provides a unified description of lattice defects and U(1) phase defects. Together, these results open a route toward first-principles modeling of defect networks in solids, metamaterials, and other ordered media.

\bibliography{cartan}
\end{document}